\documentclass{article}
\usepackage{arxiv}
\usepackage[numbers]{natbib}
\usepackage{float}
\usepackage{caption}
\usepackage{amsmath,amssymb,amsfonts} 
\usepackage{wrapfig}
\usepackage{graphicx}
\usepackage{sidecap}
\usepackage{url}
\usepackage{soul}
\usepackage{rotating}
\usepackage{xcolor}
\usepackage{multirow} 
\usepackage{booktabs}

\begin{document}
\title{Frequency-Dependent Attenuation Reconstruction with an Acoustic Reflector}   
\author{Richard Rau\thanks{\textit{rrau@ee.ethz.ch}}\ , 
Ozan Unal, 
Dieter Schweizer, 
Valery Vishnevskiy, 
Orcun Goksel
\\
\\
Computer-assisted Applications in Medicine, Computer Vision Lab, ETH Zurich, Switzerland.}
\thanks{
Funding provided by the Swiss National Science Foundation}
\maketitle

\begin{abstract}
Attenuation of ultrasound waves varies with tissue composition, hence its estimation offers great potential for tissue characterization and diagnosis and staging of pathology. 
We recently proposed a method that allows to spatially reconstruct the distribution of the overall ultrasound attenuation in tissue based on computed tomography, using reflections from a passive acoustic reflector.
This requires a standard ultrasound transducer operating in pulse-echo mode and a calibration protocol using water measurements, thus it can be implemented on conventional ultrasound systems with minor adaptations.
Herein, we extend this method by additionally estimating and imaging the frequency-dependent nature of local ultrasound attenuation for the first time.
Spatial distributions of attenuation coefficient and exponent are reconstructed, enabling an elaborate and expressive tissue-specific characterization.
With simulations, we demonstrate that our proposed method yields a low reconstruction error of 0.04\,dB/cm at 1\,MHz for attenuation coefficient and 0.08 for the frequency exponent.
With tissue-mimicking phantoms and \textit{ex-vivo} bovine muscle samples, a high reconstruction contrast as well as reproducibility are demonstrated.
Attenuation exponents of a gelatin-cellulose mixture and an \textit{ex-vivo} bovine muscle sample were found to be, respectively, 1.4 and 0.5 on average, consistently from different images of their heterogeneous compositions.
Such frequency-dependent parametrization could enable novel imaging and diagnostic techniques, as well as facilitate attenuation compensation of other ultrasound-based imaging techniques.
\end{abstract}


\section{Introduction}
Diagnostic imaging methods aim to characterize and spatially map tissue properties in order to localize and differentiate physiological and pathological structures and processes.
Conventional B-mode ultrasound images echo amplitude of ultrasound, which are (longitudinal) compressional waves, reflected from local tissue features.
There has also been several additional techniques derived for quantifying various tissue characteristics, using acoustic interaction phenomena and novel imaging sequences: 
For instance, in shear-wave elastography, focused ultrasound is used to create acoustic radiation force (a remote push) inside tissue and the local propagation speed of the resulting shear-wave is used to estimate local tissue shear-modulus~\cite{sandrin_shear_2002,eby_validation_2013,chintada_acoustoelasticity_2019}. 
For shear waves, amplitude decay over distance was measured to estimate shear-wave attenuation in~\cite{catheline2004measurement}, which was later studied for liver steatosis~\cite{barry_shear_2012}, liver transplantation~\cite{nenadic_application_2014}, and under different preloading states~\cite{chintada_nonlinear_2020}.

Compared to the above shear-wave techniques measuring shear modulus in tissue, methods determining the propagation speed of (compressional) ultrasound waves, i.e. speed-of-sound, relate to bulk modulus in tissue, and thus have a different dependency on tissue composition~\cite{glozman_method_2010}.
The estimation of speed-of-sound distribution in tissue has been studied in various setups, e.g. in transmission mode utilizing custom made transducer architectures~\cite{duric_detection_2007,malik_quantitative_2018} and in pulse-echo mode using conventional ultrasound scanners. 
For the latter, available approaches can be divided into two groups: 1)~Methods that measure apparent displacements of backscattered signals insonified from different angles~\cite{sanabria_spatial_2018, rau2019diverging,rau_ultrasound_2019}, and 2)~Methods that use an additional passive acoustic reflector to record corresponding time-of-flight values~\cite{sanabria_hand-held_2016}.
A quantification of \textit{in-vivo} speed-of-sound imaging was recently proposed for breast density~\cite{Sanabria_breast-density_18} and sarcopenia assessment~\cite{Sanabria_speed_18}.

\emph{Attenuation} of an ultrasound (US) wave is the result of several mechanisms: 
First, depending on the tissue structure dimensions, US waves may \emph{reflect} or \emph{scatter} at interfaces of different acoustic impedances.
\emph{Viscous absorption} is caused by frictious losses.
In addition, a main mode of energy loss is the result of \emph{relaxation absorption}, which is due to consecutive wave-fronts ``hitting'' the tissue that is locally recovering (bouncing back) from the push of an earlier wave-front~\cite{smith_introduction_2011}. 
These effects collectively lead to an overall \emph{ultrasound attenuation} (UA), i.e.\ amplitude decay of US signals per distance they travel.
\textit{Ex-vivo} studies of excised tissue samples~\cite{bamber_ultrasonic_1979,goss_compilation_1980} showed that UA differs between malignant and benign breast lesions~\cite{goss_comprehensive_1978} , between normal and cancerous liver~\cite{bamber_acoustic_1981b}, and between skin tissues with and without lymphedema~\cite{omura2020frequency}. 
Hence, imaging local UA-based characteristics in tissue can be a valuable diagnostic biomarker for various clinical applications.

For imaging UA, \emph{transmission} mode approaches that require complex, dedicated setups were proposed, e.g.\ with a ring transducer for scanning the breast immersed in a water bath. 
Such a method proposed in~\cite{duric_detection_2007} indicated potential for breast tumor detection and characterization.
An extension of this method later in~\cite{li_breast_2017} using a waveform tomography approach instead of a ray based approximation of acoustic propagation further improved UA imaging accuracy.
Such transmission mode setups are, however, bulky, costly, and not compatible with conventional clinical US systems with hand-held transducers, thus making UA imaging inaccessible for most clinical practice.
Imaging the attenuation of compressional ultrasound waves in \emph{pulse-echo} mode is non-trivial since one cannot observe the propagation of waves  directly using ultrasound, unlike the slower shear-waves that can be observed  using ultrasound from tissue displacements.
One method  for UA coefficient estimation utilizes adjacent frequency components of ultrasound signals as reference (spectra normalization) for cancelling out systemic effects such as focusing and time-gain-compensation, which then allows for an approximate derivation of the wave amplitude decay~\cite{gong2019system}. Another approach with hand-held devices is based on power spectrum measurements of backscattered echos over certain blocks~ \cite{coila_regularized_2018,deeba_swtv-ace:_2019,vajihi2018low,destrempes2019construction}.
Typically a reference phantom is utilized for normalization and a (spatially weighted) regularization for reconstruction~\cite{deeba_swtv-ace:_2019}.   Although indicating the potential  of UA in steatosis diagnosis~\cite{deeba_attenuation_2019}, these methods rely on the assumption that  UA in tissue  increases linearly with frequency, which however is known not always to be the case~\cite{bamber_acoustic_1981b}.

An alternative approach of estimating the UA is with a passive reflector together with a conventional US system.
An early study using such an approach for UA quantification was shown in~\cite{huang_ultrasonic_2005,chang_reconstruction_2007}, which however  required a-priori segmentation of imaged structures in order to only quantify piece-wise constant UA values in each structure.
This is then not suitable for imaging (reconstructing) arbitrary spatial distributions of UA, and due to reconstruction instabilities, only synthetic and phantom examples could be quantified, but no actual tissue samples.
Recently in~\cite{Rau_att19}, we proposed a novel reconstruction approach for arbitrary spatial UA distributions based on \emph{limited-angle computed tomography} (LA-CT) with a conventional linear-array transducer and a passive reflector.
We herein  further analyze this method and extend it to  also characterize the local frequency-dependence of UA. 
This  makes our UA imaging  applicable regardless of the frequency range,  while providing an additional novel imaging bio-marker for tissue characterization.

\section{Methods}

\begin{figure*}
\includegraphics[width=\textwidth]{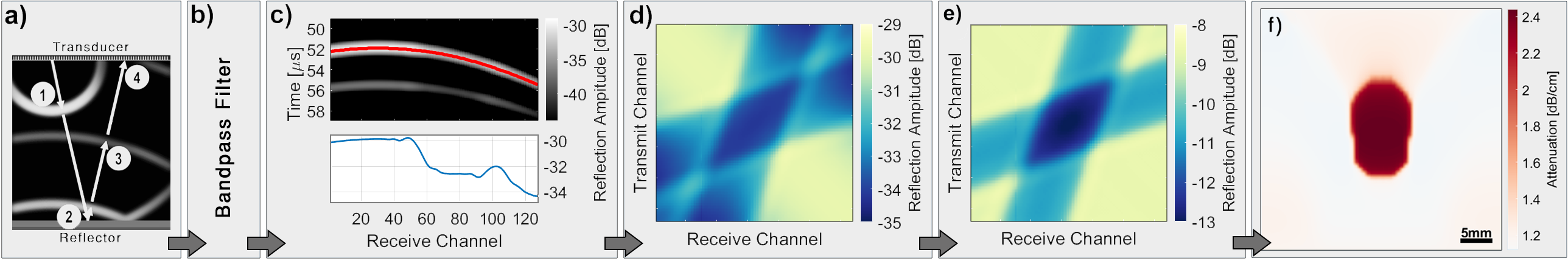}
\captionof{figure}{Processing chain for  a single band of} the proposed attenuation reconstruction with the reflector setup. \textbf{a)} Illustration of a wide-band wavefront progression in the medium and reflection back from the passive reflector. \textbf{b)} Band-limited filtering of the RF echo data  to obtain a single band. 
\textbf{c)} Echo amplitude along the detected wavefront for a single transmission. 
Reflected amplitudes for all $M$$=$$128^2$ transmit/receive channel combinations \textbf{d)} before and \textbf{e)} after calibration. \textbf{f)} Reconstructed UA. \label{fig:proc}
\end{figure*}

An overview of the proposed frequency-resolved UA reconstruction procedure is schematically depicted in Fig.~\ref{fig:proc} for a simulated example. 
Acoustic waves are emitted from a single transmit (Tx) channel and reflected from a plexiglas plate (density $\rho$=$1180\,k\mathrm{g/m^3}$, speed-of-sound $c$=$2700\,\mathrm{m/s}$), placed at a distance~$d$ from the transducer surface. 
Echo signals are recorded by all receive (Rx) elements in parallel and the process is repeated for each and every Tx element, such that a full-matrix (multistatic) echo dataset is obtained. 
A sample wide-band wavefront is illustrated in Fig.~\ref{fig:proc}a at different time-points: 1) after transmission, 2) while being reflected from the passive reflector plate, 3) during echo travel, and 4) while being recorded at the transducer. 
The wide-band echos are then filtered to obtain band-limited signals at a defined frequency spectrum centered at frequency~$f$ (Fig.~\ref{fig:proc}b).  
Subsequently, for each single element Tx event, the band-limited echo from the upper surface of the plexiglas reflector is delineated across the Rx channels (Fig.~\ref{fig:proc}c-top) and the signal envelope amplitude along this delineated reflection is derived (Fig.~\ref{fig:proc}c-bottom). 
The reflection amplitude readings for all Tx-Rx combinations (Fig.~\ref{fig:proc}d) are then calibrated (Fig.~\ref{fig:proc}e), which allows to derive the attenuation distribution in the tissue specimen by solving an inverse problem (Fig.~\ref{fig:proc}f).
The calibration and the reconstruction procedures are explained next.

\subsection{Calibration of Measurements}

 Refraction is the change in wavefront direction at interfaces with acoustic impedance variation, and diffraction is the bending of waves, e.g., around high impedance barriers~\cite{szabo_diagnostic_2013}.
We herein assume that the speed-of-sound and density, and hence acoustic impedance, in soft tissues vary marginally such that refraction and diffraction effects can be neglected.
Accordingly, band-limited ultrasound pulses at a center frequency $f$  are assumed to propagate along rays.
Note that, due to constructive and destructive interferences,  acoustic intensify from a transducer element has a response as a function of  wavefront (ray) propagation direction $\theta$, which  causes \emph{natural focusing} of energy in the main beam axis  as well as \emph{side lobe} effects.
The amplitude of the reflector  for transmitting with channel $t$ and receiving with channel $r$  can then  be described by 
\begin{equation}  \label{eq:A}
    A_{t,r} = A_{t}(\theta,f)\cdot R(\theta) \cdot S_{r}(\theta,f) \cdot \exp\bigg(\!\!-\!\!\!\int_{\text{ray}_{t,r}} \!\!\!\! \alpha(x,y,f)\ \mathrm{d} l \bigg),
\end{equation}
where $R(\theta)$ is the incident-angle dependent reflection coefficient at the reflector interface, and $A_{t}(\theta,f)$ and $S_{r}(\theta,f)$ are, respectively, the band-limited amplitude of Tx element $t$ and the sensitivity of Rx element $r$, at center frequency $f$ and transducer directivity $\theta$. The exponent describes frequency-dependent amplitude decay based on the line integral of attenuation $\alpha$ along ray$_{t,r}$ from element $t$ to $r$. 

To  estimate attenuation  $\alpha$ from $A_{t,r}$, any confounding factors in~(\ref{eq:A}), such as transducer impulse-response influence in both $A_{t}(\theta,f)$ and $S_{r}(\theta,f)$ as well as reflection characteristics in $R(\theta)$, must first be compensated for.
Considering the impulse response, the band-limited signals are normalized with a calibration experiment in water, where the speed-of-sound $c_{\mathrm{water}}$ and the attenuation coefficient $\alpha_{\mathrm{water}}$ are known from the literature, given the water temperature and imaging frequency.
By using the same predetermined frequency bandwidths and reflector distances in the water calibration as in the subsequent tissue experiments, $A_{t}(\theta,f)$ and $S_{r}(\theta,f)$ components can be effectively estimated and factored out.
However, $R(\theta)$ is still likely to differ  between any calibration and application, because of the speed-of-sound mismatch between water and tissue.
Such reflection coefficient changes at the acoustic reflector interface can nevertheless be analytically estimated using Snell's law, as the wavelength ($\approx300\,\mu\mathrm{m}$) is small compared to the reflector dimensions.
One can then write the reflection coefficient at the interface of the reflector and some medium $k$=\{water,tissue\} as follows:
\begin{equation}\label{eq:R}
    R_{k}(\theta) = \frac{m_{k}\cos(\theta)-n_{k}\sqrt{1-\frac{\sin^2(\theta)}{n_{k}^2}}}{m_{k}\cos(\theta)+n_{k}\sqrt{1-\frac{\sin^2(\theta)}{n_{k}^2}}},
\end{equation}
where speed-of-sound ratio $n_{k}$=$c_{\mathrm{reflector}}/c_{k}$ and density ratio $m_{k}$=$\rho_{\mathrm{reflector}}/\rho_{k}$.
We herein assume $\rho_{\mathrm{tissue}}$ $\approx$ $\rho_{\mathrm{water}}$ $\approx$ 1000\,kg/m$^3$ and that the speed-of-sound is dispersion-free (i.e., not frequency dependent) within the range of the utilized imaging frequency bandwidth. 
A normalized reflection amplitude matrix $B'$ shown in Fig.\,\ref{fig:proc}e can then be computed with its elements corresponding to Tx $t$ and Rx $r$ being
\begin{equation}\label{eq:normA}
    b'_{t,r} = \frac{A_{t,r,\mathrm{tissue}} R_{\mathrm{water}}(\theta)}{A_{t,r,\mathrm{water}}R_{\mathrm{tissue}}(\theta)}\,.
\end{equation}
 Substituting (\ref{eq:A}) in (\ref{eq:normA}), we can then relate calibrated and band-limited measurements at frequency $f$ to attenuation $\alpha(x,y,f)$ at image location ($x$,$y$) by
\begin{align}\label{eq:normB}
    b_{t,r} = \ln b'_{t,r} = -\int_{\text{ray}_{t,r}} \!\!\!\! \alpha(x,y,f)\ \mathrm{d}l  \approx -\!\!\!\sum_{i\in\text{ray}_{t,r}}\!\!l_{i}\alpha_{i}(f)\,,
\end{align}
where the ray integral is discretized as a summation over a reconstruction grid.
For this discretization, we use local attenuation value $\alpha(x,y)$ weighted by its path length within the grid element ($x$,$y$).

\subsection{Image Reconstruction of Attenuation}
\begin{figure*}
\centering
    \includegraphics[width=.9\textwidth]{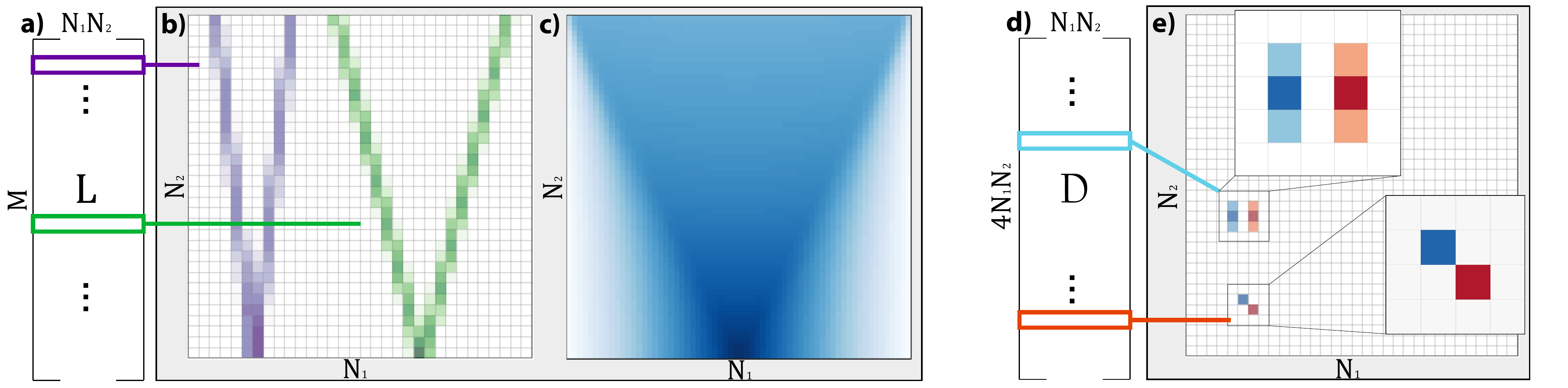}
    \caption{Reconstruction matrix $\textbf{L}$ (a) with two representative paths depicted (b). 
    The path distribution (c) illustrates the spatial coverage of tomographic projections in the imaging field of view.
    Regularization matrix $\textbf{D}$ (d) with a horizontal Sobel and a diagonal Roberts kernel (e), where blue and red indicate positive and negative components with zero-sum. 
    } \label{fig:reconst}
\end{figure*}

Given $M$ logarithms of normalized reflection amplitudes $\mathbf{b}\in\mathbb{R}^{M}$ at a specified frequency band, we perform a tomographic reconstruction of the spatial UA distribution on a  $N_1$$\times$$N_2$ spatial grid by formulating the following convex optimization problem:\hfill
\begin{equation}\label{eq:opti}
 \boldsymbol{\hat\alpha}(f) = \underset{\boldsymbol{\alpha}(f)}{\arg\min}
 \| \textbf{L}\boldsymbol{\alpha}(f)  + \mathbf{b}(f) \|_1  +  \lambda \|\textbf{D}\boldsymbol{\alpha}(f) \|_1,
\end{equation}
where $\textbf{L}\in\mathbb{R}^{M\times N_1N_2}$ is the sparse ray path matrix (cf.\ Fig.\,\ref{fig:reconst}) that implements~(\ref{eq:normB}) and $\boldsymbol{\alpha}(f)\in\mathbb{R}^{N_1N_2}$ is the reconstructed image at the center frequency $f$ of the band-pass filtered signal (cf.\ Fig.\,\ref{fig:proc}b).
Due to the ill-conditioning of $\mathbf{L}$, a regularization term controlled by weight $\lambda$ encourages spatial smoothness.
Similarly to~\cite{rau_ultrasound_2019,rau2019diverging,sanabria_hand-held_2016,sanabria_speed--sound_2018}, we herein use $\ell_1$-norm for both the data and regularization terms for robustness to outliers in, respectively, the measurements and the reconstructed image (edges)~\cite{fu_efficient_2006}.
Due to a lack of full angular coverage of measurements, regularization matrix $\mathbf{D}$ implements LA-CT specific image filtering to suppress streaking artifacts orthogonal to missing projections via anisotropic weighting of directional gradients~\cite{sanabria_hand-held_2016}. 
In axis-aligned directions a Sobel kernel, and in diagonal directions a Roberts kernel is used~\cite{rau2019diverging} (see Fig.~\ref{fig:reconst}d/e).
Similarly to~\cite{sanabria_spatial_2018}, we herein utilize a \mbox{$\kappa$$=$$0.9$} anisotropic weighting for a stronger regularization in the lateral direction, taking into account the uneven distribution of measurement paths. 
In this paper we empirically set $\lambda$=0.6 for all experiments , following~\cite{Rau_att19}.
For the numerical solution of the optimization problem~(\ref{eq:opti}), a limited-memory Broyden-Fletcher-Goldfarb-Shanno (L-BFGS) algorithm~\cite{broyden:bfgs,Fletcher:bfgs,Goldfarb:BFGS,Shanno:bfgs} from the unconstrained optimization package \texttt{minFunc}\footnote{\url{https://www.cs.ubc.ca/~schmidtm/Software/minFunc.html}} is used.

\subsection{Frequency-Dependent Attenuation Reconstruction}

\begin{figure*}
\includegraphics[width=.98\textwidth]{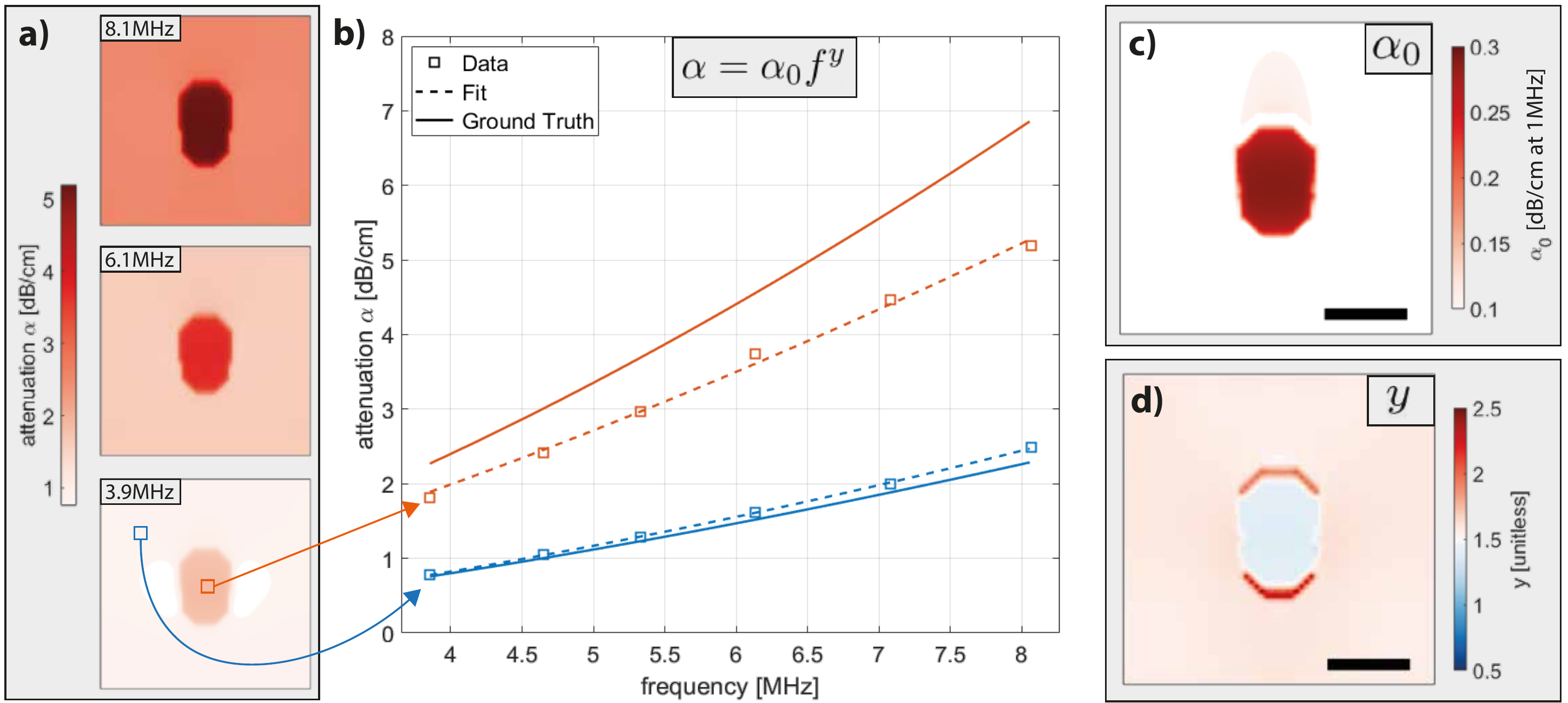}
\captionof{figure}{\textbf{a)} Attenuation reconstructions within individual frequency bands. 
\textbf{b)} Frequency-dependent attenuation characterized at two locations (orange) in and (blue) outside the inclusion.
By fitting the power-law at each pixel, the corresponding maps for attenuation coefficient $\boldsymbol\alpha_0$ (c) and the frequency exponent $\boldsymbol{y}$ (d). 
The scale-bars represent $10\,\mathrm{mm}$.
} \label{fig:freqsplit}
\end{figure*}

To characterise the frequency-dependent nature of attenuation, wide-band RF echo signals are first band-pass filtered prior to reconstruction.
In this work, for analyzing the echos acquired at an imaging center frequency of 5\,MHz, we used $1$\,MHz-bands at nominal center frequencies $[3.5, 4.5, ..., 8.5]$\,MHz. 
Band-pass filtering is herein implemented as a minimum-order filter with a stop-band attenuation of 60\,dB, using the {\fontfamily{qcr}\selectfont bandpass} function in Matlab 2019a.
For a simulated dataset, UA reconstructions at three different center frequencies are depicted in Fig.\,\ref{fig:freqsplit}a.
 The \emph{true} center frequencies after bandpass filtering are calculated as the spectral weighted average of the filtered signal, and these used for further analysis.
 Note that the true center frequencies  (e.g., $[8.1,6.1,3.9]$\,MHz as indicated in the northwest corner of each reconstruction ) are slightly  different than their nominal values  (respectively $[8.5,6.5,3.5]$\,MHz) used as bandpass centers. 
This is because the emitted wide-band pulse is approximately Gaussian with a peak at 5\,Mhz, i.e.\ more energy is concentrated around this center.
Consequently, even after a uniform filter, the resulting signals have weighted averages $f$ that are shifted towards the original peak at 5\,MHz.

In Fig.\,\ref{fig:freqsplit}b, the frequency-dependent attenuation values are shown for two regions, within the inclusion and outside.
These plots already yield a spectral fingerprint of the attenuation characteristics in each tissue.
To parametrize such spectroscopic attenuation reconstructions in a standardized way, one can assume the conventional power-law relation: 
\begin{equation}\label{eq:powerlaw}
    \alpha(f) = \alpha_0f^y\,,
\end{equation}
where attenuation is a function of frequency with medium-specific attenuation coefficient $\alpha_0$ and medium-specific exponent $y$.
Accordingly, separately at each spatial image location, $\alpha_0$ and $y$ are estimated by least-squares fitting the power-law function to the band-limited reconstructions,  yielding the parameter maps shown in Fig.\,\ref{fig:freqsplit}c and d.

Note that the attenuation term in~(\ref{eq:opti}) could have alternatively been replaced by the power-law equation~(\ref{eq:powerlaw}), for solving all frequency bandwidths simultaneously.
This could provide further robustness by regularizing reconstructions across frequencies by their expected power-law relation as well as by allowing for the direct spatial regularization of the two final parameters.
However, this would also increase the optimization problem size in several folds, impeding an efficient numerical solution.

\section{Experiments}
\noindent {\bf Metrics. } We used the following metrics for a quantitative analysis of the simulation results:
\begin{itemize}
\item Root-mean-squared-error: \\
    $\mathrm{RMSE}$ $=\sqrt{\|\hat{\boldsymbol{x}}-\boldsymbol{x}^\star\|_2^2 / N}$\,,\\
    where $\boldsymbol{x}$ represents maps of $\alpha$, $\alpha_0$, or $y$, with $N$ being the number of map pixels. The $^*$ and $\,\hat{}\,$ indicate, respectively, the ground-truth and our reconstruction.
    \item Contrast-ratio fraction: \\ 
    $\mathrm{CRF} =  \hat C/C^*$\,,\\
    where $C = 2\frac{|\mu_{\mathrm{inc}} - \mu_{\mathrm{bkg}}|}{|\mu_{\mathrm{inc}}|+|\mu_{\mathrm{bkg}}|}$ with mean inclusion and background values $\mu_{\mathrm{inc}}$ and $\mu_{\mathrm{bkg}}$, respectively, where the inclusion is delineated from the ground-truth map. 
    \item Contrast-to-noise ratio:\\
    $\mathrm{CNR} = |\mu_{\mathrm{inc}}-\mu_{\mathrm{bkg}}|/\sqrt{\sigma_{\mathrm{inc}}^2 + \sigma_{\mathrm{bkg}}^2}$\,,\\
    where $\sigma^2$ represents the variance.
\end{itemize}
\vspace{-4pt}

\begin{figure}
\centering
\includegraphics[width=.48\textwidth]{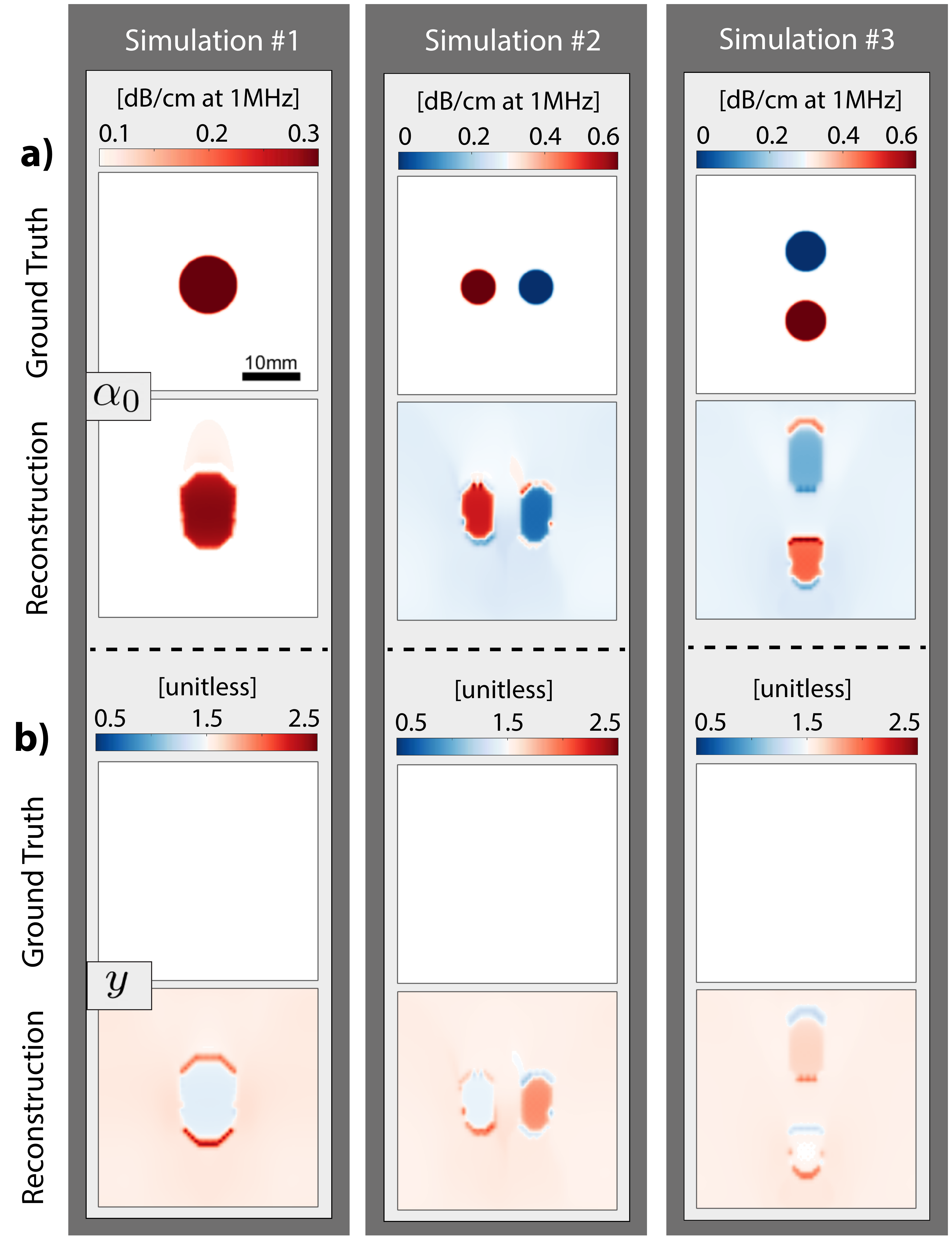}
\captionof{figure}{Frequency-dependent UA image reconstruction for three simulations.
Each column shows one simulation with the $\alpha_0$-maps (a) and the $y$-maps (b). 
Above each reconstruction map, the ground truth maps are shown. }\label{fig:sim}
\end{figure}

\vspace{1ex} \noindent {\bf Simulation Study. }
To evaluate the accuracy of frequency-dependent UA reconstructions, three simulations with different UA patterns were performed as shown in Fig.\ref{fig:sim}.  
The patterns were selected herein to assess inclusion  separability (i.e.\ resolution) along and orthogonal to the tomographic projections as well as  to assess positive and negative  contrast inclusions with respect to a background.
Acoustic propagation and received echo signals were simulated using k-Wave ultrasound simulation toolbox~\cite{treeby_k-wave:_2010} using a spatial grid resolution of $37.5\,\mu \mathrm m$.
The inclusions are simulated with a spatial variation in attenuation constant $\alpha_0(x,y)$.
As this toolbox does not allow to spatially vary the power-law dependence parameter $y$, this value is kept constant at $y=1.5$ for this numerical evaluation. 
Full-matrix acquisition was simulated at a center frequency of $5\,\mathrm{MHz}$ with a $5$-cycle pulse. 
The transducer was simulated as a  linear array with 128 channels (i.e.\ $M = 128^2$) with a pitch of $300\,\mu \mathrm m$.

\vspace{1ex} \noindent {\bf Phantom and \textit{Ex-vivo} Tissue Study. }
A tissue-mimicking gelatin phantom was prepared with 10\% gelatin in water per weight. 
We created non-scattering and scattering phantoms; for the latter, using 1\% per-weight Sigmacell Cellulose Type 50 (Sigma Aldrich, St. Louis, MO, USA). 
 Typical for the assessment of ultrasound imaging modalities, such gelatin phantoms mimic homogeneous soft tissues with a speed-of-sound of approximately 1540\,m/s, thus representative e.g.\ of the liver for the scattering phantom and the blood for the scatter-free one.
Fresh bovine skeletal muscle was used as \textit{ex-vivo} tissue samples.
Using  different combinations of the above, we created four phantoms  listed in Table~\ref{tab:pht}.
\begin{table}\centering
\caption{\textit{Ex-vivo} muscle -- gelatin combinations for experiments.}
\label{tab:pht}
\begin{tabular}{|@{ }l@{ }|@{ }l@{ }|@{ }l@{ }|}  \hline
{\bf Pht.} & {\bf Inclusion} & {\bf Background} \\ \hline
\#1 & single \textit{ex-vivo} muscle & scattering gelatin  \\ \hline
\#2 & single \textit{ex-vivo} muscle & non-scattering gelatin \\ \hline
\#3 & single scattering gelatin  & \textit{ex-vivo} muscle \\ \hline
\#4 & two non-scattering gelatin & \textit{ex-vivo} muscle  \\ \hline
\end{tabular}
\end{table}
Experiments were performed with the samples submerged in distilled water at room temperature.
Muscle samples were imaged with fibers oriented orthogonal to the imaging plane, to prevent potential confounding effects from anisotropic acoustic propagation along tissue fibers, as this is currently not accounted for in our model. 
Data acquisition was performed with a Verasonics Vantage 128-channel system connected to a Philips ATL L7-4 transducer (Verasonics, Kirkland, WA, USA). 
Analogous to the simulation setup, we used a  wideband Tx  with a center frequency of $5\,\mathrm{MHz}$ and a pulse length of $5$ cycles.
The reflector distance varied from 30\,mm to 46\,mm depending on the sample size. 
For normalization of the \textit{ex-vivo}/phantom amplitude matrices at a given reflector depth $d$, we acquired calibration measurements a-priori in distilled water at room temperature at multiple reflector depths \{31, 35, ...,47\}\,mm.
For a given phantom measurement, we then computed the desired calibration values $b_{t,r}$ at $d$ via linear interpolation from the closest two calibration  depths.

\section{Results and Discussion}

\subsection{Simulation Experiments}

Three simulations (\#1--\#3) shown in Fig.~\ref{fig:sim} were conducted (case $\#$1 is the one illustrated earlier in Fig.~\ref{fig:freqsplit}), as representative cases with inclusions separated laterally and axially, as well with both positive and negative contrast to background. 
Given the ground truth maps in Fig.~\ref{fig:sim} it can be seen that the attenuation coefficient $\alpha_0$ is successfully reconstructed in all cases. 
It is observed that with increasing UA distribution complexity (i.e., from simulation \#1 to \#3), the reconstruction accuracy decreases. 
In particular, axially-separated inclusions are reconstructed poorer compared to those laterally separated. 
This is due to the transmit-receive paths traversing the tissue in a mostly axial direction and hence lateral projections being missing in the resulting LA-CT, as was also reported and discussed in the context of speed-of-sound reconstruction in~\cite{Rau_att19,sanabria_hand-held_2016}.
In general, within the inclusions a shift of $\alpha_0$ towards background values is observed.
This is due to the lack of lateral projections combined with the regularization term in (\ref{eq:opti}) axially elongating the inclusions, hence spreading them over a larger area.
Since the cumulative UA effect shall stay the same, thus per-pixel inclusion values are effectively averaged with the background, inversely proportional to such artifactual area increase.  

 In Fig.~\ref{fig:sim} the frequency exponent $y$-maps should not present any contrast, since the k-Wave simulated ground-truth data does not allow to spatially vary~$y$. 
Reconstructed $y$-maps are seen to exhibit a deviation from such constant ground-truth value, with a fixed background offset and some artifacts along the axial edges of inclusions.
The latter artifact originates from the artificial axial elongation of inclusions due to above-described LA-CT nature. 
This  herein manifests itself stronger with increased frequency band, i.e. the higher the frequency band, the longer an inclusion axially is  in the reconstruction.
This effect can be observed in Fig.~\ref{fig:freqsplit}a, where the inclusion shape at $8.1$\,MHz is axially slightly longer compared to that at $3.9$\,MHz. 
This artifactually higher UA values around the axial edges at higher frequencies are then  erroneously attributed by the model to increased $y$ values  around these edges. 
 Note that an inverse problem with multiple-bands formulated in the same linear system would allow to constrain and regularize the reconstructions at the parametric level (i.e.\ edges enforced consistently at all frequencies) and hence may prevent such axial edge artifacts in $y$-maps.

To quantify the performance of the frequency-dependent UA reconstructions, we evaluate RMSE, CNR, and CRF for each simulation and report in Tab.\ref{tab:sim_freqdep} the average values and standard deviations over all three simulations. 
RMSE of $\alpha_0$ is $0.04\pm0.02$~dB/cm at 1\,MHz.
 Since ground-truth $\alpha_0$ values reach up to $0.6$~dB/cm at 1\,MHz, this demonstrates that an accurate characterization of the attenuation coefficient is possible.
With an average RMSE of $0.08\pm0.02$ for $y$, the relative estimation error with respect to the prescribed $y=1.5$ is below 6\%, indicating a relatively good estimation of frequency exponent. 
The contrast metrics (CNR and CRF) for $\alpha_0$ demonstrate that the inclusions can be successfully distinguished in $\alpha_0$ reconstructions, with a contrast similar to their prescribed ground-truth value. 
Note that the contrast metrics are only computed for $\alpha_0$, since no contrast  was prescribed for $y$ in the simulations. 
\renewcommand{\arraystretch}{1}
\begin{table}
  \centering
  \caption{Quantitative evaluation of frequency-dependent UA reconstructions in simulations.}
    \begin{tabular}{l|c|c}\label{tab:sim_freqdep}
                & $\alpha_0$                        &  $y$  \\  \hline
        RMSE    & $0.04\pm0.02$~dB/cm at 1\,MHz   & $0.08\pm0.02$    \\ \hline
        CNR     & $0.4\pm0.2$  & -  \\ \hline
        CRF     & $71\pm28~\%$                     & -  \\ \hline
    \end{tabular}
\end{table}

\begin{figure}
\centering
\includegraphics[width=.48\textwidth]{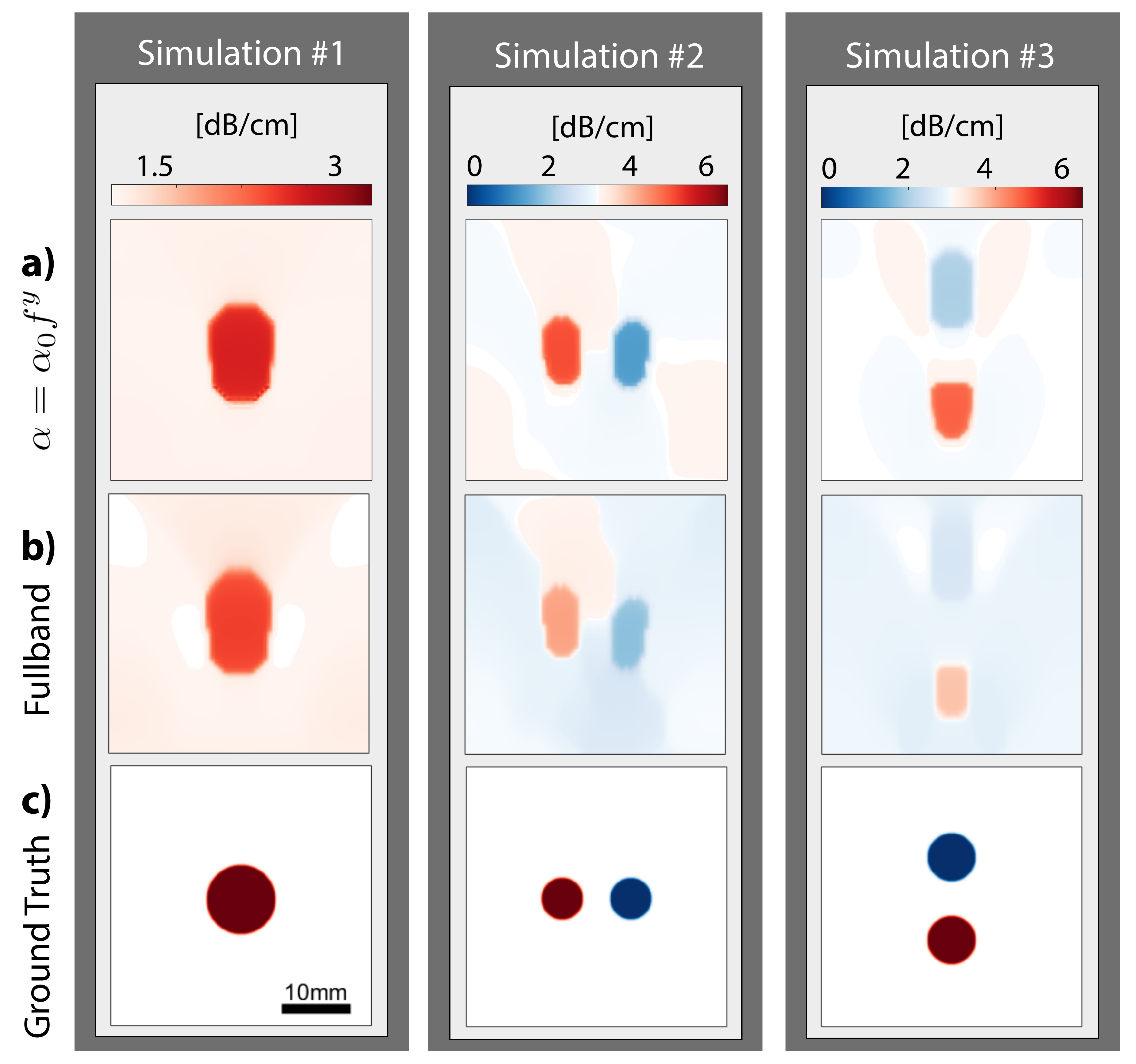}
\captionof{figure}{Comparison of frequency-dependent and fullband UA imaging. 
The columns represent different simulations with the attenuation images derived (a)~by the frequency-dependent estimations using (\ref{eq:powerlaw}) and (b)~by the fullband as in~\cite{Rau_att19}. 
Ground-truth attenuation maps in (c) are derived using the ground truth $\alpha_0$ and $y$ given the transmit center frequency.  }\label{fig:frequency_formula_comparison}
\end{figure}

To estimate how frequency-dependent UA imaging compares to the frequency-averaged fullband UA reconstructions \cite{Rau_att19}, we compute the overall attenuation $\alpha$-maps using $\alpha = \alpha_0f^y$ for each pixel at the known  (average) transmit center frequency of 5~MHz (Fig.~\ref{fig:frequency_formula_comparison}a). 
The corresponding fullband reconstructions are shown in Fig.~\ref{fig:frequency_formula_comparison}b along with the ground-truth images derived at 5\,MHz from $\alpha_0$- and $y$-maps prescribed in the corresponding simulations (also shown in Fig.~\ref{fig:frequency_formula_comparison}c). 
It is seen that estimating $\alpha_0$ and $y$  separately, i.e.\ without their effects combined in the overall attenuation, improves UA characterization  substantially.
This observation is corroborated by  all metrics reported in Tab.\ref{tab:sim_overallatt}  comparatively for both methods,
\renewcommand{\arraystretch}{1.4}
\begin{table}
  \centering
  \caption{ Comparison of attenuation imaging using frequency-dependent estimation or fullband frequency averaging.  Better result for each metric and simulation case is highlighted in bold.}
    \begin{tabular}{p{.1mm}cl|r|r|r|}
                 &              & \textbf{Simulation} & \textbf{\#1}   & \textbf{\#2}   & \textbf{\#3} \\
    \midrule
    \multirow{2}[4]{*}{\begin{sideways}\textbf{\,\,\,\,\,RMSE}\end{sideways}} & \multirow{2}[4]{*}{\rotatebox{90}{\quad\footnotesize[dB/cm]}} & $\alpha_0f^y$ & \textbf{0.28}         & \textbf{0.4}          & \textbf{0.58} \\ 
\cline{3-6}                 &              & Fullband     & 0.31         & 0.52         & 0.72 \\
    \midrule
    \multirow{2}[4]{*}{\begin{sideways}\,\,\,\,\,\textbf{CNR}\end{sideways}} & \multirow{2}[4]{*}{\rotatebox{90}{\quad\footnotesize[unitless]}} & $\alpha_0f^y$ & \textbf{5.3}          & \textbf{5.7}          & \textbf{1.5} \\
\cline{3-6}                 &              & Fullband     & 4.0            & 2.9          & 0.3 \\
    \midrule
    \multirow{2}[4]{*}{\begin{sideways}\,\,\,\,\,\textbf{CRF}\end{sideways}} & \multirow{2}[4]{*}{\rotatebox{90}{\quad\footnotesize[\%]}} & $\alpha_0f^y$ & \textbf{77.2}         & \textbf{51.8}         & \textbf{31.1} \\
\cline{3-6}                 &              & Fullband     & 68.8         & 34.6         & 11.2 \\
    \bottomrule
    \end{tabular}%
  \label{tab:sim_overallatt}%
\end{table}%
 where an improvement is observed in every simulated case for every metric. 
The average improvement in UA estimation accuracy (RMSE) is 17\%, with a substantial enhancement in reconstruction contrast: an average CNR improvement of 176\% and an average CRF improvement of 80\%.
In an actual imaging scenario such improved contrast would enable easier identification and characterization of abnormality, especially in challenging cases where more heterogeneous compositions are imaged (as in simulations $\#2$ and $\#3$).
 In comparison to a single value of frequency-averaged reconstructions, frequency-dependent UA imaging  yields two biomarkers,  which can either be used in independently  to compensate confounding effects or in combination  as a multi-parametric imaging biomarker, 
both of which can thus be potentially more expressive for  tissue characterization. 

\subsection{\textit{Ex-vivo} Experiments} 

\begin{figure}
\centering
\includegraphics[width=.49\textwidth]{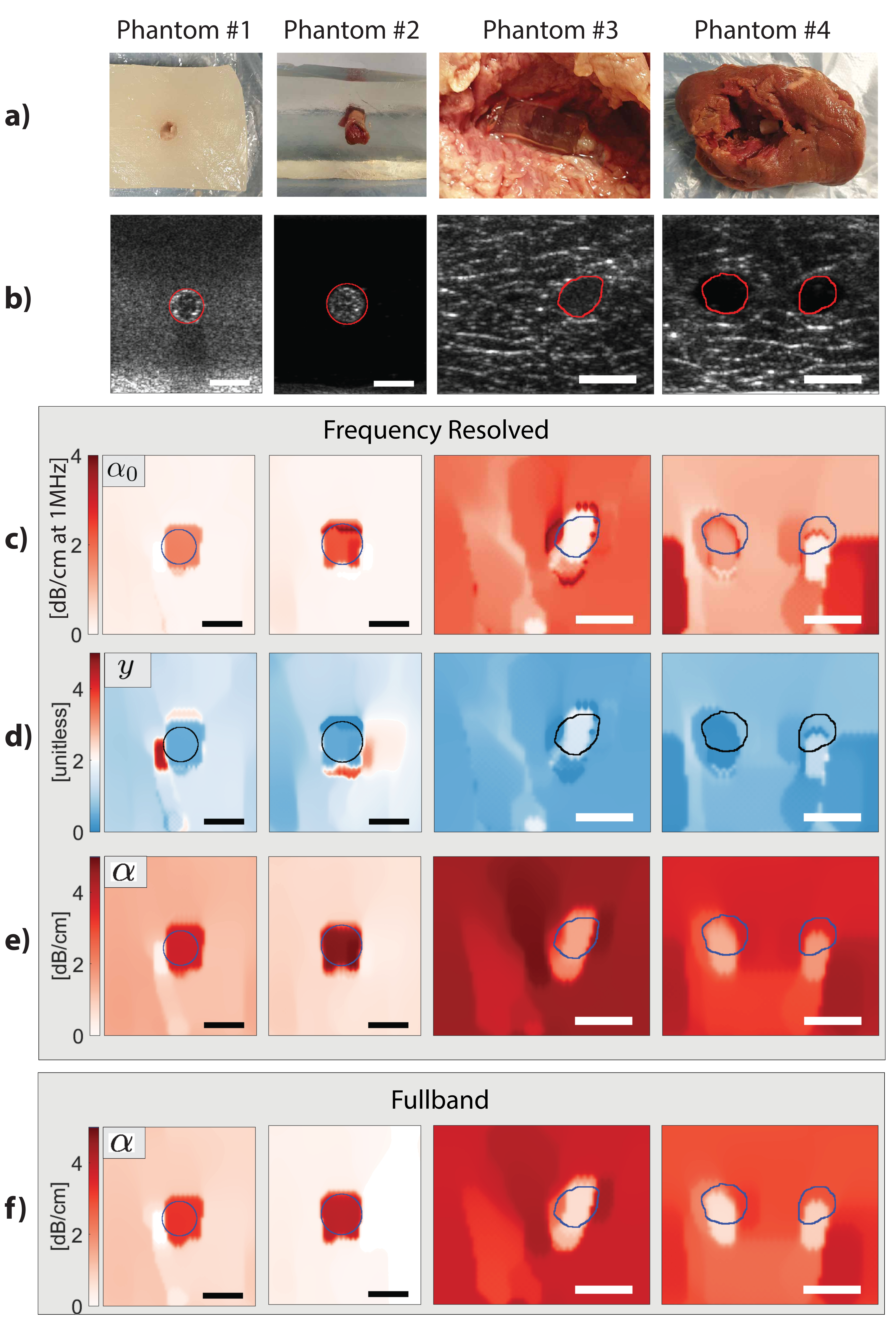}
\captionof{figure}{Reconstructions from gelatin phantoms and \textit{ex-vivo} bovine muscle. \textbf{a)} Sample pictures of muscle inclusions in gelatin ($\#$1) with and ($\#$2) without cellulose; and muscle samples with gelatin inclusions ($\#$3) with and ($\#$4) without cellulose.
\textbf{b)} Corresponding B-Mode images.
Reconstructions of \textbf{d)} attenuation coefficient $\alpha_0$ and \textbf{e)} exponent $y$ using the proposed frequency-resolved method.
\textbf{f)} Frequency-averaged reconstructions using the fullband spectrum~\cite{Rau_att19}. 
The scale-bars represent $10\,\mathrm{mm}$.}
\label{fig:exvivo}
\end{figure}

The results of the gelatin phantom and  \textit{ex-vivo} experiments are shown in Fig.\ref{fig:exvivo} 
with their corresponding B-Mode images and attenuation reconstructions for the frequency-dependent ($\alpha_0$, $y$) parameters as well as the fullband frequency-averaged ($\alpha$) attenuation (as in~\cite{Rau_att19}).
For the frequency-resolved approach, an equivalent $\alpha$ is also computed (for comparison to~\cite{Rau_att19}) based on (\ref{eq:powerlaw}) using the reconstructed coefficient and exponent maps. 
For the phantoms \#1 \&~\#2 with muscle inclusions in gelatin, a clear contrast is visible for all UA reconstructions (Fig.\ref{fig:exvivo}c-f).
With our proposed frequency-dependent method, it can be seen that the observed attenuation contrast originates from a variation not only in attenuation coefficient but also in frequency-dependence. 
For phantom \#3, the reconstructed values with our proposed frequency-resolved method are consistent with phantoms \#1 and \#2, where the inclusion and background are inverted.
Compared to the fullband method, $\alpha_0$- and $y$-maps are seen to better preserve inclusion geometry with a reduced axial elongation, corroborating the earlier findings in the simulations. 
This  geometric accuracy is especially important in cases, where the inclusions are not directly observable in the B-Mode images, e.g.\ in phantom~\#3 where the scattering gelatin and the surrounding muscle  being isoechoic. 

To suppress axial edge artifacts, especially in $y$-maps, a possible solution would be to first delineate target regions for UA characterization and then to solve the reconstruction problem with UA constrained as piece-wise constant within each such predefined region. 
Such regions could be delineated manually or automatically from B-Mode images or from $\alpha_0$-maps of initial unconstrained UA reconstructions.
 It is also possible to reconstruct all frequency bands in a single linear system for consistent regularization and an implicit model-fitting across frequencies (e.g., to better leverage higher SNR frequencies), although this would yield much larger inverse-problems for reconstruction.

In general, the background reconstructions  are seen to deteriorate towards the two lower corners, likely due to fewer {tomographic projections of} these regions, as also illustrated in Fig.~\ref{fig:reconst}c.
 Such errors may  also propagate to the rest of the image,  e.g.\ deteriorating how well the inclusions, especially those nearby the lower corners, are reconstructed.
Consequently, phantom~\#4 with two inclusions on the sides  yields a poorer reconstruction.
Indeed, in contrast to the superior performance of the frequency-resolved method for phantoms \#1--\#3, for phantom \#4 it yields poorer results compared to the fullband, with localization and geometry estimation errors.
Surprisingly, the two inclusions with the same material are not even characterized similarly in the reconstructions of neither $\alpha_0$ nor~$y$. 
Nevertheless, similar to the fullband representation, an attenuation map $\alpha$ (Fig.\ref{fig:exvivo}e) can be derived from the frequency-dependent $\alpha_0$- and $y$-maps using (\ref{eq:powerlaw}), which is then seen to recover any lost contrast, even for phantom \#4.  
This observation is supported quantitatively by average CNR of attenuation across  \textit{ex-vivo} samples being comparable as $(3.3\pm2.6)$ for the fullband method and $(3.1\pm2.2)$ using (\ref{eq:powerlaw}) from frequency-resolved parameter maps.
These results also demonstrate that both our frequency-resolved and fullband methods can find applicability in  different clinical scenarios.

\begin{figure}
\centering
\includegraphics[width=.49\textwidth]{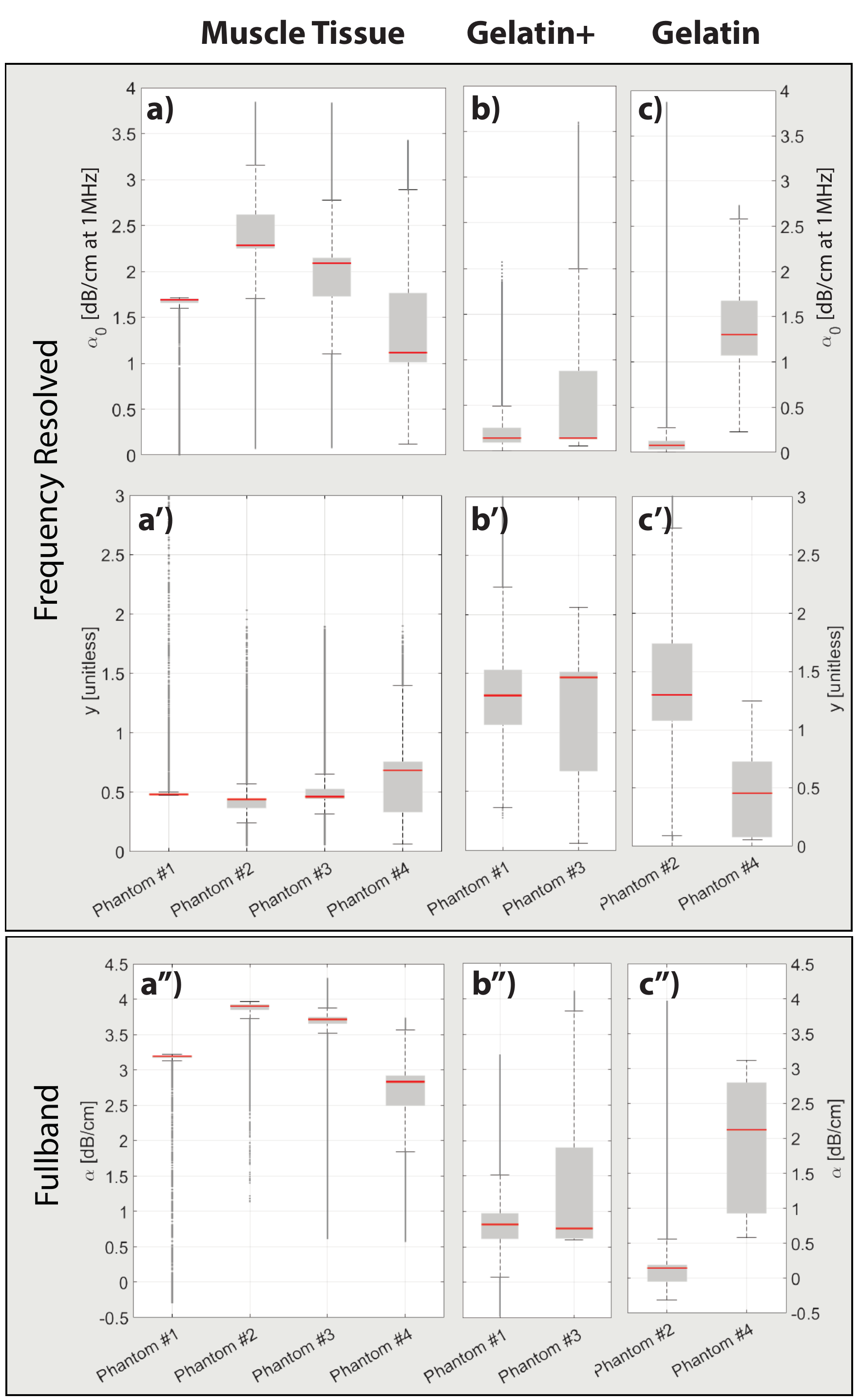}
\captionof{figure}{Distributions of attenuation values for different materials: muscle tissue (a), gelatin with cellulose (b), and pure gelatin (c).
Frequency-dependent parameters $\alpha_0$ and $y$ are shown, respectively, in (a-c) and (a'-c').
Fullband frequency-averaged attenuation is shown in (a''-c'').}
\label{fig:exvivo_quant}
\end{figure}

To quantify reproducibility of our attenuation imaging and characterization for the same material in different configurations, we conducted the following analysis on the \textit{ex-vivo} dataset:
Since three materials in total (bovine muscle tissue, pure gelatin, and gelatin-cellulose mixture) were used in phantoms \#1-\#4, 
it is expected and desired that the same material is reconstructed as the same value across different phantoms. 
To quantify such consistency and constancy, in Fig.\ref{fig:exvivo_quant} %
we compare distributions of  per-pixel reconstructions within manually-delineated material regions depicted in Fig.\ref{fig:exvivo}.
We present such distributions along with the median values (red lines in Fig.\ref{fig:exvivo_quant}) for frequency-dependent $\alpha_0$ and $y$ reconstructions as well as for fullband $\alpha$ reconstructions.
In this figure, $\alpha$ derivation using (\ref{eq:powerlaw}) is omitted since it closely follows fullband $\alpha$ distributions, as demonstrated earlier in Fig.\ref{fig:exvivo}.
To quantify constancy across the phantoms, we compare the median values, where their statistical distribution (mean and standard-deviation) for the same material across phantoms is reported in Table~\ref{tab:exphant}. 
\renewcommand{\arraystretch}{1.2}
\begin{table}
  \centering
  \caption{Reproducibility of material characterization over different phantoms for the two methods.}
    \begin{tabular}{l|c}\label{tab:exphant}
                        & $\alpha$\ \ (Fullband)   \\  \hline
        Muscle tissue   & $(3.41 \pm 0.49)$\,dB/cm      \\ \hline
        Gelatin+        & $(0.79 \pm 0.04)$\,dB/cm      \\ \hline
        Gelatin         & $(1.13 \pm 1.40)$\,dB/cm      \\ \hline
                        &         \\ 
                        & $\alpha_0f^y$\ \ (Frequency-resolved)  \\ \hline 
        Muscle Tissue   &   $(1.86 \pm 0.51)$\,dB/(MHz$^{(0.49 \pm 0.11)}$cm)   \\ \hline
        Gelatin+        &   $(0.15 \pm 0.00)$\,dB/(MHz$^{(1.36 \pm 0.12)}$cm)   \\ \hline
        Gelatin         &   $(0.71 \pm 0.88)$\,dB/(MHz$^{(0.83 \pm 0.51)}$cm)
    \end{tabular}
\end{table}
%
For the scattering gelatin with cellulose (referred as \textit{gelatin+}), a low standard-deviation in Table~\ref{tab:exphant} indicates a high reproducibility considering all three UA parameters (also can be observed in Fig.\ref{fig:exvivo_quant}b/b'/b"). 
Despite an expected natural tissue heterogeneity, all reconstructions of the muscular tissue also follow relatively tight distributions with somewhat similar values as seen in Figs.\ref{fig:exvivo_quant}a/a'/a". 
Muscle tissue appears to have a high $\alpha$ and 
$\alpha_0$ whereas a low exponent $y$, and
accordingly presents a stark contrast to gelatin+. 
For the pure gelatin case, we expect an overall attenuation lower than gleatin+ due the lack of scattering that contributes to attenuation.
Indeed, for phantom \#2 compared to \#1 \& \#3, both $\alpha_0$ and $\alpha$ are significantly lower, as expected, cf.\ Fig.\ref{fig:exvivo_quant}b/b''/c/c''.
Interestingly, $y$ is similar with and without scattering, cf.\ Fig.\ref{fig:exvivo_quant}b'/c'.
This can be explained by the fact that frequency-dependent attenuation is mostly due to relaxation and viscous absorption, and not due to scattering. 
Also the cellulose scatterers of size 50\,$\mu$m are smaller than the wavelength of the highest frequency band, thus interacting with all involved frequency bands similarly as a scattering source.
Such observations unfortunately cannot be made from phantom \#4, the reconstructions of which as mentioned above were affected severely by artifacts.

\begin{SCfigure}
\centering
\includegraphics[width=.24\textwidth]{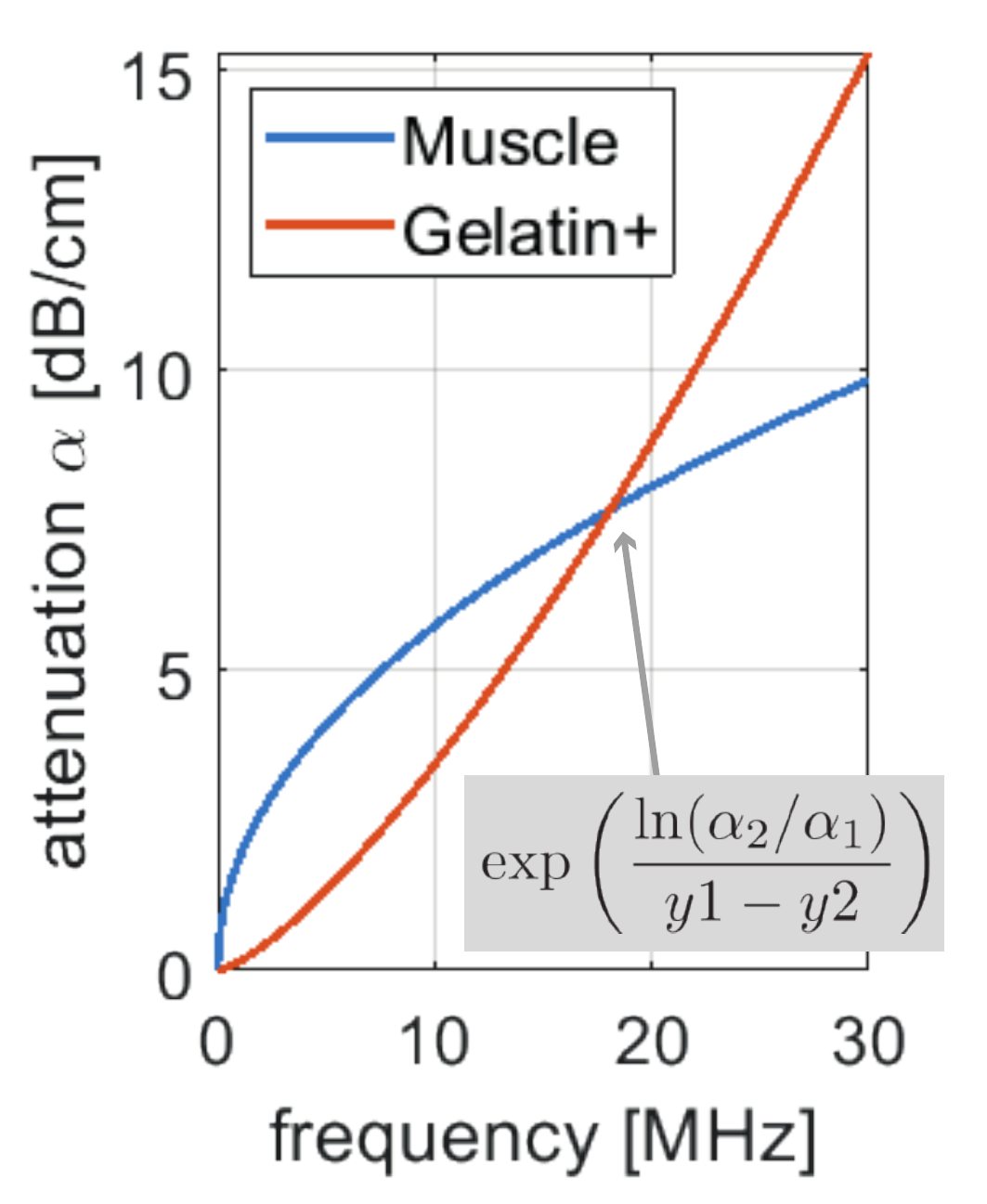}
\captionof{figure}{Extrapolated attenuation of the muscle and gelatin+ materials. 
At $\approx$18\,MHz, calculated as shown in the figure, no discrimination  between the two materials would theoretically be possible.
\vspace{1.5cm}}
\label{fig:attoverlap}
\end{SCfigure}

\subsection{Extrapolation of Attenuation Characteristics}

With the separate characterization of the power-law coefficients of frequency-dependent attenuation, observed attenuations can now be extrapolated to a wider hypothetical frequency range using the power-law relation~(\ref{eq:powerlaw}).
This is presented in Fig.~\ref{fig:attoverlap} for the muscle and gelatin+ materials using the average values reported in Tab.~\ref{tab:exphant}. 
From Fig.~\ref{fig:attoverlap} we can conclude that  within typical medical ultrasound bandwidth, i.e.\ under 20\,MHz, contrast between the attenuation $\alpha$ of these two materials would be maximized at $5.6$\,MHz as $\Delta\alpha=2.8$\,dB/cm. 
Conversely, at $\approx18$\,MHz as seen in Fig.~\ref{fig:attoverlap}, the two materials will likely not exhibit any contrast at all.
This demonstrates that using a frequency-averaged UA imaging, the achieved contrast may highly depend on the imaging frequency.
Similarly, an imaging frequency unfit for a practical scenario may potentially lead to suboptimal tissue characterization and even potential misclassification, e.g., of a tumor. 
Our proposed frequency resolved attenuation imaging method redeems such limitation of having to select an optimal imaging frequency, which is likely unknown \textit{a-priori}, by allowing for a characterization independent of  utilized frequency range (and thus imaging transducer and  frequency setting). 

\subsection{Further Discussion}

Herein we only take the speed-of-sound difference between water calibration and target tissue into account for compensating for the reflection coefficient differences, cf.~(\ref{eq:normA}). 
Tissue speed-of-sound difference  between the water calibration  and target tissue may, however,  also affect the angular beam profile of the transducer elements, hence confounding the assumption of amplitude matrix constancy between water and tissue in~(\ref{eq:normA}). 
Nevertheless, we believe such discrepancy of the beam profiles to be minimal, given the successful reconstruction results demonstrated with both our frequency-resolved and fullband attenuation imaging methods.

In our experiments we utilized a transmit pulse of five cycles.
This could be shortened for the full spectral bandwidth of the transmit pulse to be extended up to the full transducer bandwidth, hence allowing for a wider range of reconstruction frequencies.
 For our purposes, within a given transducer bandwidth, either the echos from several narrowband Tx's can be used to separately analyze the data or a single wideband Tx can be analyzed in subbands with post-processing as presented herein. Our method is thus applicable using existing US transducers and systems, provided access to raw RF data.

Similarly to most earlier works in US-CT, we herein assume a straight ray  wavefront propagation, which is however only true for a constant speed-of-sound distribution. 
Local speed-of-sound inhomogeneities may indeed refract and aberrate wavefronts, resulting in inaccurate amplitude readings as well as the assumed wavepaths (e.g.\ Fig.\ref{fig:reconst}b) being different than actual wavepaths -- both  possibly leading to inaccuracies in reconstruction.
A potential remedy would be to correct  any ray refractions based on  an a-priori speed-of-sound reconstruction, e.g.\ using time-of-flight measurements as in~\cite{sanabria_speed--sound_2018}. 

 With the proposed frequency dependent UA imaging method, the opposite side of an imaged tissue needs to be accessible with a passive reflector, which limits potential anatomical regions to the breast, the extremities, male genitelia, etc.  
These are similar to the locations applicable for passive reflector based speed-of-sound imaging as well as for full US-CT methods. 
It may further be possible to use a passive reflector during open or laparoscopic surgeries, such as using a reflector with a so-called ``pick-up'' US transducer~\cite{Schneider_tracked_2015}.

For a multi-parametric tissue characterization, our proposed method add s two additional imaging biomarkers complementary to robust speed-of-sound imaging methods~\cite{rau2019diverging,rau_ultrasound_2019}, which  is potentially superior to elastography~\cite{glozman_method_2010}.
 Hence, compared to existing ultrasound imaging modalities, a more comprehensive assessment of tissue composition and potential pathology is enabled using the proposed frequency-dependent attenuation imaging.

In addition to using the proposed attenuation imaging for diagnostic purposes, the derived UA distribution can also be utilized to improve other ultrasound imaging modalities. 
For instance, similarly to~\cite{rau_ultrasound_2019}, where speed-of-sound imaging is used to correct for aberration effects in beamforming, UA  maps can be used to compensate  signal gain individually for each Rx-element and beamformed pixel combination.
This can homogenize B-mode image appearance by removing artificial contrasts, can enhance optoacoustic~\cite{mercep_transmissionreflection_2019} and functional imaging~\cite{rau_3d_2018}, and can further improve displacement estimations, e.g.\ in shear wave elastography.

A practical limitation of the herein proposed frequency-resolved attenuation imaging is the computational time required for the reconstructions due to the iterative numerical optimization employed. 
A deep variational network solution such as in~\cite{vishnevskiy_deep_2019,vishnevskiy2018image} with reconstruction times in the range of milliseconds could allow to overcome this restriction towards real-time imaging.

\section{Conclusions}
In this paper we have presented a novel approach for reconstructing frequency-dependent ultrasound attenuation distribution in tissue, which allows to locally characterize tissue with two biomarkers -- namely the attenuation coefficient and the frequency exponent.
This method allows for a more accurate reconstruction of UA distribution compared to our earlier method in~\cite{Rau_att19} that reconstructs fullband frequency-averaged value of a single attenuation parameter. 
Our experiments with numerical simulations, tissue-mimicking gelatin phantoms, and \textit{ex-vivo} tissue samples indicate a high overall reconstruction accuracy with a mean RMSE of simulated phantoms being 0.04\,dB/cm at 1\,MHz for attenuation coefficient $\alpha_0$ and 0.08 for frequency exponent $y$.
Compared to fullband reconstructions, the frequency-dependent method offers a  substantial average improvement  of up to 176\% in the analyzed metrics, RMSE, CNR and CRF, as reported in Table~\ref{tab:sim_overallatt}.
Requiring only a minimal hardware add-on as a passive acoustic reflector, the proposed imaging is readily translatable to clinical applications where the opposite side of the imaged tissue region is accessible, e.g.\ imaging the breast, the extremities, and male genitalia.

\bibliographystyle{elsarticle-harv}
\bibliography{refs}

\end{document}